\documentclass[twocolumn,showpacs,preprintnumbers,amssymb]{revtex4}

\usepackage{graphicx}
\usepackage{epsfig}
\usepackage{dcolumn}
\usepackage{bm}
\def\gsim{\lower0.5ex\hbox{$\:\buildrel >\over\sim\:$}}
\def\lsim{\lower0.5ex\hbox{$\:\buildrel <\over\sim\:$}}

\begin{document}

\preprint{BNL-HET-06/19}

\title{Signature of heavy Majorana neutrinos at a linear collider:
Enhanced charged Higgs pair production}

\author{David Atwood$^a$}%
\email{atwood@iastate.edu}
\author{Shaouly Bar-Shalom$^b$}
\email{shaouly@physics.technion.ac.il}
\author{Amarjit Soni$^c$}%
\email{soni@bnl.gov}
\affiliation{$^a$Department of Physics and Astronomy, Iowa State University,
Ames, IA 50011, USA\\
$^b$Physics Department, Technion-Institute of Technology, Haifa 32000, Israel\\
$^c$Theory Group, Brookhaven National Laboratory, Upton, NY 11973, USA}

\date{\today}

\begin{abstract}
A charged Higgs pair can be produced at an $ee$ collider through a
t-channel exchange of a heavy neutrino ($N$) via $e^+ e^- \to H^+
H^-$ and, if N is a Majorana particle, also via the lepton number
violating (LNV) like-sign reaction $e^\pm e^\pm \to H^\pm H^\pm$.
Assuming no a-priori relation between the effective $eNH^+$ coupling
($\xi$) and light neutrino masses, we show that this interaction
vertex can give a striking enhancement to these charged Higgs pair
production processes. In particular, the LNV $H^-H^-$ signal can
probe $N$ at the ILC in the mass range $100 ~{\rm GeV} \lsim m_N
\lsim 10^4 ~{\rm TeV}$ and with the effective mixing angle, $\xi$,
in the range $10^{-4} \lsim \xi^2 \lsim 10^{-8}$ - well within its
perturbative unitarity bound and the $\beta \beta_{0\nu}$ limit. The lepton number conserving (LNC)
$e^+ e^- \to H^+ H^-$ mode can be sensitive to, e.g., an ${\cal O}(10)$ TeV heavy Majorana neutrino at a 500 GeV International Linear Collider (ILC), if $\xi^2 \gsim 0.001$.
\end{abstract}


\maketitle

The discovery of neutrino oscillations, which indicates mixing
between massive neutrinos, stands as the first direct
evidence for physics beyond the Standard Model (SM). Thus, the SM, or for that matter also the Minimal Supersymmetric Standard Model (MSSM) and any other multi-Higgs doublet model, has to be
expanded to include massive neutrinos that
mix. Since there is still no understanding of the nature of these
massive neutrinos, {\it i.e.}, Majorana or Dirac-like, the extensions to these
models can go either way. However, from the theoretical point of view,
the more natural and therefore more appealing
way is to generate sub-eV Majorana neutrinos
by adding heavy right-handed neutrino singlet fields ($N$ with a mass
$m_N$) and relying on the seesaw mechanism
which yields $m_\nu \sim m_{D}^2/m_N$, where $m_\nu$ is the solar and atmospheric neutrino masses and $m_D$ are the Dirac mass terms.
Thus, the classic seesaw mechanism links light neutrino masses
($m_\nu \sim 10^{-2}$ eV) with new physics ($m_N$) at the GUT scale (if $m_D \sim m_W$) or at the multi-TeV scale (if $m_D \sim m_e$) - both are well motivated theoretically. In either case, this minimal seesaw setup yields a vanishingly small heavy-to-light neutrino mixing: $U_{\ell N} \propto \sqrt{m_\nu/m_N}$, in which case heavy neutrinos effectively decouple.

In this work we will assume no a-priori relation between the
couplings and light neutrino masses. Indeed, interesting models have
been proposed, which assume that the underlying
physics for sub-eV neutrino masses goes beyond the standard, minimal,
seesaw framework (see e.g., \cite{beyond1,beyond2,ma,beyondss}), thus allowing the mixing angle
$U_{\ell N}$ to depart from its classical seesaw value (i.e., $\sqrt{m_\nu/m_N}$).
Our primary purpose in this paper is to propose collider
experiments that can sensitively probe the relevant masses and mixing
angles.

The fact that a Majorana mass term violates lepton number by two
units, {\it i.e.} $\Delta L = \pm 2$, has striking phenomenological
implications, some of which have been extensively studied in the past few decades -
the most noticeable one being neutrinoless double beta decay $(A,Z)
\to (A,Z+2) + e^- + e^-$ \cite{2beta}. With the upcoming CERN Large Hadron Collider (LHC) and the planned high-energy $ee$ machines, the study of LNV
collider signals in processes involving heavy Majorana neutrinos is
particularly well motivated. Indeed, such high-energy LNV signals have been extensively studied
in the past two decades: at $pp$ and $p{\bar p}$ collisions
\cite{delAguila,oldpp,ali1,Wpapers,han}, at an $e^+ e^-$ \cite{delAguila,buch1},
$e^-e^-$ \cite{rizzo} and $e^- \gamma$ collisions
\cite{pilaftsis1,delAguila} and at an $e p$ machine
\cite{delAguila,ali2,buch1}. In addition, LNV mediated by exchanges of a heavy Majorana N can be manifest in top-quark and W-boson decays \cite{ourtWpaper} and in rare charged meson decays \cite{ali1}.

In this letter we propose that charged Higgs pair production at an $ee$ machine like the ILC (International Linear Collider) can serve as a very sensitive probe of heavy neutrinos (N) with masses up to $m_N \sim 10^4$ TeV. The two processes we will explore are:

\begin{eqnarray}
&& e^+ e^- \to H^+ H^- \label{pm}~, \\
&& e^\pm e^\pm \to H^\pm H^\pm \label{mm}~.
\end{eqnarray}

\noindent These processes are generated at the tree level via a t-channel (and u-channel in the case of the like-sign $H^\pm H^\pm$ signal) exchange of a heavy $N$, which can have a sizable coupling to $e-H^+$ when $m_N >> m_W$ (see below). Note also
that, while the $H^+ H^-$ signal is not sensitive to the nature of N, i.e., be it Majorana or Dirac particle, the like-sign $H^\pm H^\pm$ signal is generated only if N is a Majorana neutrino.

\begin{figure*}[htb]
\epsfig{file=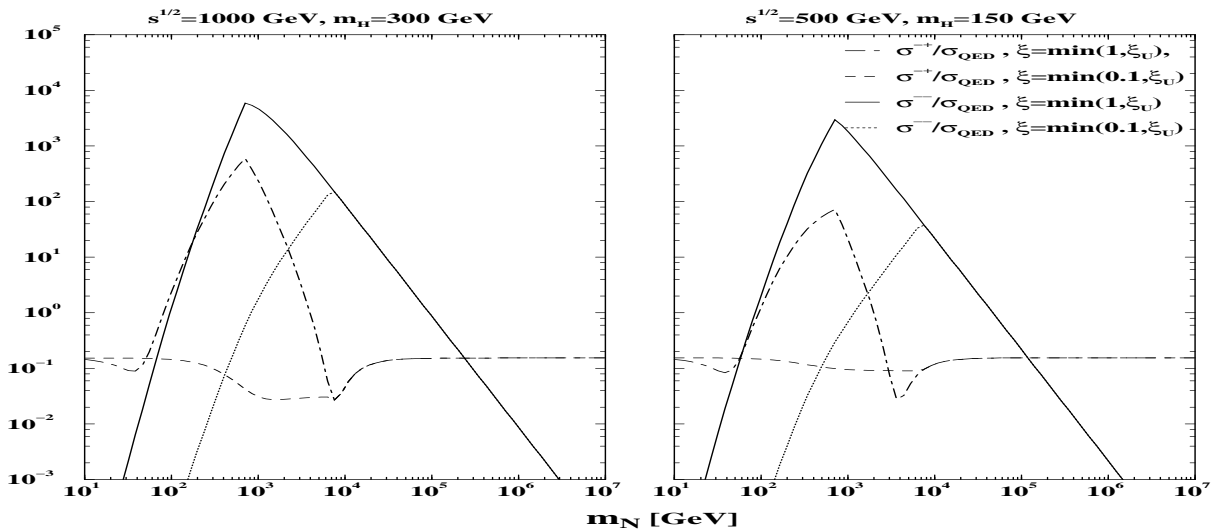,height=16cm,width=7cm,angle=270}
\caption{\emph{The LNV and LNC cross-sections $\sigma^{--}$ and $\sigma^{-+}$ in units of $\sigma_{QED} \equiv \sigma(e^+e^- \to \gamma^\star \to \mu^+ \mu^-)$, as a function of $m_N$ and with $\xi={\rm min}(0.1,\xi_U)$ or $\xi={\rm min}(1,\xi_U)$,
for a 500 GeV (right figure) and 1000 GeV (left figure) $ee$
collider with charged Higgs masses
$m_H = 150$ and 300 GeV, respectively.}}
\label{figsmm}
\end{figure*}

In any model with more than one Higgs doublet, a right handed neutrino (be it Majorana or Dirac-like) can couple to a charged Higgs via its Yukawa interactions with the scalar doublets $Y L \Phi_i N$. The $H^+ \ell N$ Yukawa-like coupling can, therefore, be written as:

\begin{equation}
{\cal L}_{H^+ \ell N} = \frac{g}{\sqrt{2}} \bar N
\left(f_v^\ell \frac{m_\ell}{m_W} R + f_v^N \frac{m_D}{m_W} L \right) \ell H^+ + h.c. \label{YHln0}~,
\end{equation}

\noindent where $m_D$ and $m_\ell$ are the neutrinos and charged leptons Dirac mass terms, respectively, and the $f_v$'s are some functions of the Higgs doublets VEV's. For example, $f_v^\ell=(f_v^N)^{-1}=\tan\beta$ within an extended supersymmetric or a non-supersymmetric type II two Higgs doublet model with heavy right handed neutrinos.

Since we will be interested in the effects of a heavy Majorana-type neutrino, we assume $N$ to have a Majorana mass term of the form $M N N/2$. Thus, we rewrite the $H^+ e N$ interaction Lagrangian (\ref{YHln0}) in a generic form, in terms of the physical mass of $N$, $m_N$, and assuming $f_v^e m_e << f_v^N m_D$:

\begin{equation}
 {\cal L}_{H^+ e N} = \frac{g}{2 \sqrt{2}} \xi  \frac{m_N}{m_W} \bar N (1-\gamma_5) e H^+ + h.c. \label{YHln}~,
\end{equation}

\noindent where the dimensionless parameter $\xi$ defines an effective $H^+ e N$ coupling strength that will be the key factor which controls the size of the cross-section for reactions (\ref{pm}) and (\ref{mm}). In particular, we absorb in $\xi$ any relation between $m_D$ and $m_N$ which will depend on the mechanism responsible for the generation of neutrino masses. Depending on the size of $\xi$, the interaction term in (\ref{YHln}) can have striking implications for $N-H^+$ phenomenology at an high-energy $ee$ collider, in particular, for the charged Higgs pair production mechanisms in (\ref{pm}) and (\ref{mm}).

As mentioned earlier, we will assume no a-priori relation between $m_N$ and light neutrino masses and mixing, thus, adopting a model independent approach towards the effective $H^+ e N$ coupling $\xi$, and treating it as a free parameter (assuming for simplicity that it is real). As such, $\xi$ should be subject only to existing experimental constraints and perturbative unitarity if applicable (see below).
However, since there is no direct experimental constraint on $\xi$ that we know of,
we will use the seesaw-like framework as a guide for estimating the experimentally allowed size of $\xi$. In particular, within a seesaw-like mechanism for generating
light neutrino masses, the Dirac mass $m_D$ is related to the physical heavy Majorana mass by $m_D \sim U_{\ell N} m_N$, where $U_{\ell N} \sim m_D/m_N$ being the heavy-to-light neutrino mixing angle (see e.g., Pilaftsis in \cite{beyondss}).
Thus, in this case we obtain (see Eqs.~\ref{YHln0} and \ref{YHln}):

\begin{equation}
 \xi \equiv f_v^N \cdot U_{eN} \label{xi}  ~.
\end{equation}

\noindent It should be emphasized again that, although the minimal seesaw framework restricts the mixing $U_{\ell N}$ to be vanishingly small,
one cannot rule out $U_{\ell N} \sim {\cal O}(1)$ even for $m_N \sim {\cal O}(100)$ GeV, based on seesaw-like scenarios which go beyond the minimal framework, such as radiative seesaw, double seesaw and fine-tuned relations or symmetries between the mass matrices $m_D$ and $m_N$, as was noted for example in \cite{beyond1,beyond2,ma,beyondss}.

Using the seesaw-like motivated relation in (\ref{xi}), the existing experimental constraints on $U_{eN}$ can be applied as an estimate for the allowed size of $\xi$.
In particular, $U_{eN}$ is constrained from precision electroweak data (e.g., from invisible Z-decays) \cite{kagan}:
$|U_{eN}|^2 \leq 0.012$ at $90\%$ CL if $m_N > m_Z$. This does not rule out
$\xi \sim {\cal O}(1)$, since $f_v^N$ which depends on the details of the scalar
sector of the underlying theory, can be of order 10.
A stronger constraint on $U_{eN}$ comes from neutrinoless double-$\beta$ decay ($\beta \beta_{0 \nu}$) \cite{limit2}:
$\sum_N U_{eN}^2/m_N \leq 5 \cdot 10^{-5}~{\rm TeV}^{-1}$. We note though, that the
$\beta \beta_{0 \nu}$ bound, although tighter, is subject to cancelations if there are more than one heavy N's that couple to an electron, in particular if they have different masses and mixings (see e.g., discussion by G. Belanger {\it et al.} in \cite{rizzo}).

As noted above, another constraint can be applied on the $\xi - m_N$ plane from perturbative unitarity by demanding that $\Gamma_N < m_N/2$. In particular, in our simplified framework
$N$ can decay via:
$N \to W^\mp \ell^\pm ,~Z \nu_\ell,~H \nu_\ell,~H^\mp \ell^\pm $,
where the decay rates into the gauge-bosons is $\propto |U_{eN}|^2$, see e.g., \cite{pilaftsis1}. Taking $M_N^2 >> m_W^2, m_Z^2, m_{H^0}^2, m_{H^+}^2$, the total decay width of $N$ (i.e., summing over the above decay channels) is:

\begin{equation}
 \Gamma_N \approx 3 |\xi|^2  \Gamma_N^0 \cdot \left( 1 + \frac{1}{|f_v^N|^{2}} \right) ; ~\Gamma_N^0 \equiv \frac{g^2}{64 \pi} \frac{m_N^2}{m_W^2} m_N ~.
\end{equation}

\noindent Thus, taking $|f_v^N|^{2} >> 1$ (i.e., assuming that the total N width is dominated by its decays to the Higgs-bosons), perturbative unitarity imposes the following bound:$^{[1]}$\footnotetext[1]{We note that partial wave unitarity of $e^-e^- \to H^- H^-$ also sets a bound on $\xi$ which is, however, weaker than $\xi_U$ coming from
$\Gamma_N < m_N/2$. We have also checked that the measured $g-2$ of the electron does not further constrain $\xi$.}
\begin{equation}
 \xi \lsim \xi_U \sim \frac{4}{g} \sqrt{\frac{2 \pi}{3}} \cdot \frac{m_W}{m_N} \sim
0.715 \frac{TeV}{m_N} \label{ubound}~.
\end{equation}

\noindent This gives e.g., $\xi \lsim 1$ for $m_N \gsim 700$ GeV or $\xi \lsim 0.1$ for $m_N \gsim 7$ TeV.
In what follows, we will always apply the unitarity bound in (\ref{ubound}), or restrict $\xi$ to be smaller than either 1 or 0.1 when the unitarity bound does not apply.

The cross-section for the LNV $e^- e^- \to H^- H^-$ mode is given by:

\begin{equation}
 \sigma^{--}=\sigma_{QED} \cdot \xi^4 \frac{3 \beta}{32 s_W^4}  \frac{r_N^3}{r_W^2}
\left( \frac{2}{w^2-\beta^2} + \frac{\ln\left(\frac{w+\beta}{w-\beta}\right)}{\beta w}
 \right) ~,
\end{equation}

\noindent where $\sigma_{QED} \equiv \sigma(e^+ e^- \to \gamma^\star \to \mu^+ \mu^-) = 4 \pi \alpha^2/3/s$, $s_W \equiv \sin\theta_W$, $\beta \equiv \sqrt{1-4 r_H}$, $w \equiv 1+2r_N-2r_H$ and
$r_x \equiv m_x^2/s$. Thus, in the limit $m_N^2 >> s$, we obtain
$\sigma^{--} \propto \xi^4 m_N^2 / m_W^4$, i.e., the cross-section grows with $m_N$ for a constant effective mixing $\xi$. This growth behavior is, however,
``cured'' when we impose the perturbative unitarity bound in (\ref{ubound}):
\begin{equation}
 \sigma^{--}(m_N^2 >> s) |_{\xi=\xi_U} \sim \frac{32 \pi \beta}{9}
\frac{1}{m_N^2} \label{bigN}~.
\end{equation}
In Fig.~\ref{figsmm} we plot the LNV cross-section $\sigma^{--}$ in units of
$\sigma_{QED}$, as a function of $m_N$
for both a 500 and 1000 GeV $ee$ machines. For illustration, we take
$m_{H^+} = 150$ and 300 GeV, for the
$\sqrt{s}=500$ and 1000 GeV colliders, respectively.
We consider the cases
$\xi = 0.1$ and $\xi=1$, imposing the unitarity bound such that
$\xi={\rm min}(0.1,\xi_U)$ and $\xi={\rm min}(1,\xi_U)$, respectively
(recall that the unitarity bound implies $\xi_U \sim 0.1$
already for $m_N \sim 7$ TeV). As expected, there is a sharp decrease in $\sigma^{--}$ which occurs once $\xi$ enters
the unitarity bound regime ($\xi \to \xi_U$), for which
the cross-section drops as $m_N^{-2}$ as seen in (\ref{bigN}).

\begin{figure}[htb]
\epsfig{file=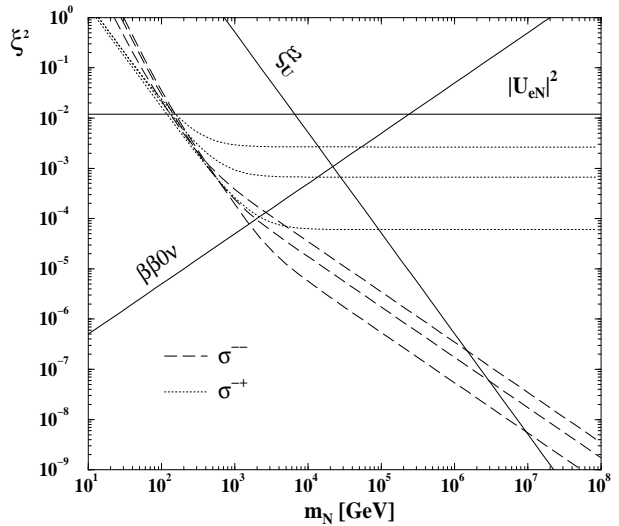,height=7cm,width=8cm}
\caption{\emph{Discovery limit for $e^-e^- \to H^-H^-$ (long-dashed lines)
and for $e^-e^+ \to H^-H^+$ (dotted lines), at
an ILC with $\sqrt{s}=500$ GeV (upper line), $\sqrt{s}=1$ TeV (middle line)
and $\sqrt{s}=3$ TeV (lower line), see also text.
The parameter space above the lines
corresponds to the observable events.
Also plotted (solid lines):
the experimental limits on $U_{eN}^2$ from $\beta \beta_{0\nu}$ and
on $|U_{eN}|^2$ from
invisible Z-decays, as well as the perturbative unitarity bound from (\ref{ubound}).
Here the parameter space above the lines are ruled out.}}
\label{figlimits}
\end{figure}

In the mass range $100 ~{\rm GeV} \lsim m_N \lsim {\rm few} \cdot 10^3 ~{\rm TeV}$,
despite fixing $\xi = {\rm min} (0.1,\xi_U)$ due to unitarity, we still see that
$\sigma^{--} \sim 10^{-3} \cdot \sigma_{QED}$.
Thus, recalling that any given design of an $ee$ machine relies on having 10000 events per unit $\sigma_{QED}$, one expects more than 10 $e^- e^- \to H^- H^-$ events even if
$m_N \sim 10^3-10^4$ TeV.
In Fig.~\ref{figlimits} we plot a ``naive'' discovery limit
(demanding 10 events for discovery)
for $e^-e^- \to H^-H^-$
in the $m_N - \xi^2$ plane,
at an ILC with $\sqrt{s}=500$ GeV, 1 TeV and 3 TeV, setting $m_H = 150,~300$ and 600 GeV, respectively, and with an integrated luminosity
of $L=100 \times (\sqrt{s}/{\rm TeV})^2$ fb$^{-1}$ which corresponds to
$L \cdot \sigma_{QED} \sim 10000$ at each c.m. energy.
We see that the like-sign Higgs pair production signal
 is indeed striking at an ILC; it
is potentially observed even if $m_N \sim 10^4$ TeV and
$\xi^2 \sim 10^{-8}$ - an extremely small value for the effective
neutrino mixing angle which is
consistent with the unitarity bound and
more than 7
orders of magnitudes smaller than
the $\beta \beta _{0\nu}$ bound at $m_N \sim 10^4$ TeV!

Let us now turn to the lepton number conserving (LNC) process $e^+ e^- \to H^+ H^-$.
There are 3 diagrams contributing to this process: the two ``standard'' (i.e.,
within multi-Higgs models) s-channel $\gamma$ and $Z$ exchanges and a 3rd t-channel N exchange one.
The cross-section $\sigma^{-+} \equiv \sigma(e^-e^+ \to H^- H^+)$ can be decomposed accordingly as $\sigma^{-+} \equiv \sigma^{-+}_S + \sigma^{-+}_{SN} + \sigma^{-+}_N$, where the subscripts $S,~N$ and $SN$ denote the standard s-channel $\gamma,Z$ contribution, the t-channel $N$ contribution and the interference between the standard and the $N$-exchange diagrams. While $\sigma_S$ can be found in the literature \cite{sigmas}, the new $N$ exchange contributions to the total cross-section are:
\begin{eqnarray}
  && \sigma^{-+}_{SN} =  C_N \frac{1- 4 r_Z s_W^2 c_W^2}{c_W^2 (1-r_Z)}
\left[  \ln\left(\frac{w+\beta}{w-\beta}\right) -\frac{\beta w}{2 v} \right]
~, \nonumber \\
 && \sigma^{-+}_N = C_N \frac{2 \xi^2}{w^2-\beta^2} \frac{r_N}{r_W}
\left[ w \ln\left(\frac{w+\beta}{w-\beta}\right) -2 \beta \right] \label{hmhpcsx}~,
\end{eqnarray}
\noindent where $v \equiv r_N + (1-w)^2/4$ and
\begin{equation}
 C_N \equiv \sigma_{QED} \cdot \xi^2 \frac{3v}{32 s_W^4} \frac{r_N}{r_W} ~.
\end{equation}

For a constant $\xi$ and in the limit $m_N^2 >>s$, both $\sigma_{SN}^{-+}$ and $\sigma_N^{-+}$ approach a constant with respect to $m_N$.
As $\xi$ enters the
perturbative unitarity bound regime, i.e., $\xi \to \xi_U \sim 1/m_N$,
the interference term dominates since
$\sigma_{SN}^{-+}/\sigma_N^{-+} \propto m_N^2/s$, but
still drops with $m_N$ as $\sigma_{SN}^{-+} \propto 1/m_N^2$.
Thus, in this large $N$ mass range the total $H^+H^-$ cross-section converges to the standard value: $\sigma^{-+}(m_N^2 >>s,\xi = \xi_U) \to \sigma_{S}^{-+}$. This behavior can be seen in Fig.~\ref{figsmm}, where we have plotted
$\sigma^{-+}$ in units of
the point cross-section $\sigma_{QED}$, as a function of $m_N$
for both a 500 and 1000 GeV $ee$ machines, for which we take
$m_{H^+} = 150$ and 300 GeV, respectively.
As in the LNV $H^-H^-$ case, we consider
$\xi = 0.1$ and $\xi=1$, imposing the unitarity bound such that
$\xi={\rm min}(0.1,\xi_U)$ and $\xi={\rm min}(1,\xi_U)$.

Note that around $m_N \sim \sqrt{s}$ the t-channel N exchange
in $H^+H^-$ production can become sizable. We, thus, define
the number of standard deviations ($N_{SD}^N$) with which the t-channel $N$ exchange in $e^- e^+ \to H^- H^+$ can be detected at an ILC:
\begin{equation}
N_{SD}^N = \frac{\sqrt{L}(\sigma^{-+}-\sigma_S^{-+})}{\sqrt{\sigma^{-+}}} \label{nsd} ~.
\end{equation}

In Fig.~\ref{figlimits} we plot the $5\sigma$ ($N_{SD}^N=5$)
discovery contours of the $N$-exchange contribution to
$e^-e^+ \to H^- H^+$ in the $\xi^2 - m_N$ plane. As in the $H^-H^-$ case, we take an ILC with $\sqrt{s}=500$ GeV, 1 TeV and 3 TeV (setting $m_H = 150,~300$ and 600 GeV, respectively), and an integrated luminosity of $L=100 \times (\sqrt{s}/{\rm TeV})^2$ fb$^{-1}$.
Although not as promising as the LNV $H^-H^-$ signal,
we see that the $N$-mediated $H^- H^+$ LNC signal can be detected at $5\sigma$ in an ILC with e.g., $\sqrt{s}=1(3)$ TeV, for a very heavy $N$
with a mass in the range $10(2)~{\rm TeV} \lsim m_N \lsim 30(100)~{\rm TeV}$ and
$\xi^2 \sim 6 \cdot 10^{-4}(6 \cdot 10^{-5})$ - a value well within the $\beta \beta _{0\nu}$ bound and consistent with the unitarity bound.

\begin{figure}
\epsfig{file=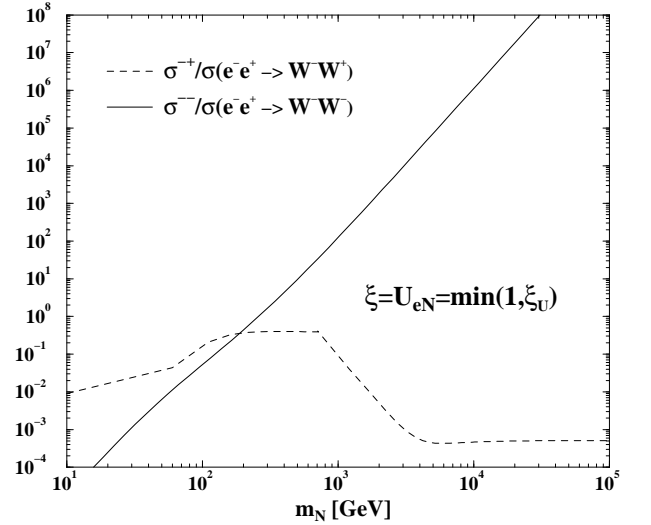,height=7cm,width=8cm}
\caption{\emph{The cross-sections ratios
$\sigma^{--}/\sigma(e^-e^- \to W^-W^-)$ and $\sigma^{-+}/\sigma(e^-e^+ \to W^-W^+)$, as a function of $m_N$, for
$\xi = U_{eN} ={\rm min}(1,\xi_U)$ at a 500 GeV ILC.}}
\label{figWW}
\end{figure}

Admittedly, our criteria for a $5\sigma$ discovery (or 10 events in the $H^-H^-$ case), as plotted in Fig.~\ref{figlimits},
is rather naive since we do not take into account the possible charged Higgs decays, their branching ratios and the corresponding detection efficiencies.
Nonetheless, this simplified discovery criteria allows
one to get an idea of the rather striking prospects of these charged Higgs pair production signals (especially the LNV $H^-H^-$ one), as potential probes of heavy Majorana neutrinos.
In particular, to further appreciate the
significance of both the LNV and LNC
Higgs pair production signals,
we plot in Fig.~\ref{figWW} the ratios
$\sigma^{--}/\sigma(e^-e^- \to W^-W^-)$ and
$\sigma^{-+}/\sigma(e^-e^+ \to W^-W^+)$ at a 500 GeV
ILC,$^{[2],[3]}$
as a function of $m_N$
and for $\xi = U_{eN} ={\rm min}(1,\xi_U)$, i.e., assuming $f_v^N=1$.

Take for example the $H^-H^-$ signal for which we get
$\sigma^{--}/\sigma(e^-e^- \to W^-W^-) > 10$ already for $m_N \gsim \sqrt{s}$ (=500 GeV).
Recall that $\sigma(e^-e^- \to W^-W^-) \propto U_{eN}^4$ is the inverse $\beta \beta_{0\nu}$ (t-channel N exchange) process, which was extensively studied as a potential discovery channel for a heavy $N$ at an ILC \cite{rizzo}, and which stands as
the leading background to the $e^- e^- \to H^-H^-$ process. Thus, the $H^-H^-$ signal
should be easily detected at a 500 GeV ILC if $m_N \gsim 500$ GeV.
Clearly, if $m_H > m_t +m_b$, then the $H^-H^-$ signal can be searched for via $e^-e^- \to H^-H^- \to bb \bar t \bar t $. In this case the effective
$H^-H^-/W^-W^-$ ratio is further (dramatically) enhanced by a $(2 \to 2)/(2\to4)$ phase-space factor.
As for the LNC $H^-H^+$ channel, we get $\sigma^{-+}/\sigma(e^-e^+ \to W^-W^+) \gsim 0.1$ for $100 ~{\rm GeV} \lsim m_N \lsim 1000 ~{\rm GeV}$. Thus, here also, in particular if $m_H > m_t +m_b$, the N-exchange contribution in the $H^-H^+$ channel should be easily accessible
to a 500 GeV ILC if $m_N \sim {\cal O}(1)$ TeV, even if $\xi^2 < 10^{-2}$ -
see also Fig.~\ref{figlimits}.

To summarize, we have examined the potential role of the charged Higgs pair production
processes $e^-e^+ \to H^-H^+$ and $e^-e^- \to H^-H^-$ at an ILC, as possible
probes of new dynamics mediated by heavy neutrinos, N. We have shown that the $m_N/m_W$
enhancement factor in the $e-N-H^+$ interaction vertex (i.e., compared to the $e-N-W^+$ vertex) makes these processes extremely sensitive to the t-channel $N$-exchange that drives them. In particular, we find that the LNV $H^-H^-$ signal
is sensitive to a heavy neutrino with a mass up to $10^4$ TeV and an effective Yukawa-like coupling to an electron as small as $\xi \sim 10^{-4}$.
For example, in a supersymmetric non-minimal seesaw-like framework with heavy Majorana neutrinos, for which $\xi = U_{eN} \cdot f_v^N$ and $f_v^N = 1/\tan\beta$, one expects
$\xi = U_{eN}/\tan\beta$, where $U_{eN}$ is the $\nu_e-N$ mixing angle. Thus, taking
$\tan\beta \sim {\cal O}(30)$ or $\tan\beta \sim {\cal O}(3)$, the LNV
$H^-H^-$ signal can probe a very heavy $N$ with $m_N \sim 10^4$ TeV for
the very small values of the neutrino mixing angles $U_{eN}^2 \sim 10^{-5}$ or
$U_{eN}^2 \sim 10^{-7}$, respectively.
This striking result is compatible with both the $\beta \beta_{0\nu}$ and the perturbative unitarity
bounds on the $U_{eN}^2 - m_N$ plane and, to the best of our knowledge,
stands out as an exceptional signal of
heavy Majorana neutrinos at large colliders.

In the case of the LNC signal $e^+e^- \to H^+H^-$, we find that
an ILC with a center of mass energy of 1 TeV, can potentially probe a heavy Majorana
with a mass $m_N \sim {\cal O}(10)$ TeV, if $\xi^2 \sim 0.001$ - also within the
$\beta \beta_{0\nu}$ and the perturbative unitarity
bounds on the $U_{eN}^2 - m_N$ plane.

Finally, in passing, we note that recently an interesting model
that can be tested at the ILC through our charged
Higgs pair production mechanisms was suggested by Ma in \cite{ma}.
In Ma's model a second scalar doublet $(\eta^+, \eta^0)$ is introduced and the light neutrinos acquire sub-eV masses through a loop involving both $N$ and $\eta^0$ - a radiative seesaw mechanism. As a consequance, the
coupling $e-N-\eta^+$ is much less constrained (than in the minimal seesaw scenario)
and should, therefore, be studied at the ILC via the reaction $e^- e^- \to \eta^- \eta^-$, where each of the $\eta^-$'s in the final state can further decay via $\eta^- \to \eta^0 W^-$, $\eta^0$ being the SU(2) partner of $\eta^+$.
It is also worth noting that, in this model, $\eta^0$ can be the lightest stable particle due to an exact $Z_2$ symmetry of the model, making it a good candidate for dark matter \cite{ma}. In such a case, the reaction $e^- e^- \to \eta^- \eta^-$ will lead to a unique signal of two same-charge $W$-bosons and a missing energy carried by the two $\eta^0$'s.

We thank Gad Eilam, Tao Han, Ernest Ma and Jose Wudka for very useful discussions.
S.B.S thanks the hospitality of the
theory group in Brookhaven National Laboratory where part of
this study was performed.
This work was supported in part by US DOE Contract Nos.
DE-FG02-94ER40817 (ISU) and DE-AC02-98CH10886 (BNL).

\footnotetext[2]{We have included the contribution from the t-channel
N-exchange diagram in $\sigma(e^-e^+ \to W^-W^+)$.}
\footnotetext[3]{Following suggestions by
Tao Han, we have also considered the constraints on $U_{eN}$ from
the LEP2 measurements of $\sigma(e^-e^+ \to W^-W^+)$ \cite{lep2}.
We found that $e^-e^+ \to W^-W^+$  is not sensitive at LEP2 to the
N-exchange diagram if $U_{eN} \lsim 0.1$, i.e., within its
bound from invisible Z decays.}

\end{document}